
\baselineskip=22pt
\magnification=\magstep1
\font\germ=eufm10
\def\sl{{\hbox{\germ sl}}\,}
{\rm \hfill
December 6,\ 1992\break}
\medskip
{\bf
\centerline{Free field realization of $q$-deformed primary fields for
$U_q(\widehat{\sl}_2)$}}
\medskip\medskip
\centerline{Atsushi MATSUO}\medskip
\centerline{Department of Mathematics}\par\noindent
\centerline{Nagoya University, Nagoya 464-01, Japan}
\medskip\medskip\noindent
\baselineskip=17pt plus 3pt

\def\parn{\par\noindent}
\def\medn{\medskip\medskip\noindent}
\def\med{\medskip\medskip}

\def\bar{\overline}
\def\QED{{\it \hfill Q.E.D.}}
\def\>{{\bf \rangle}}
\def\<{{\bf \langle}}
\def\C{{\bf C}}
\def\Z{{\bf Z}}
\def\End{{\rm End}}
\def\ker{{\rm ker}}
\def\U{U_q}
\def\ge{{\hbox{\germ g}}\,}
\def\qqquad{\qquad\qquad}

\def\and{\hbox{ \  and \ }}
\def\ds{\displaystyle }
\def\no{\leqno\ \ }

\def\vbn{\vfill\break\noindent}

\def\F{{F}}
\def\FF{{\cal F}}

\def\Res{{\mathop{\rm Res}}}
\def\a{\alpha}
\def\b{\beta}
\def\d{\delta}
\def\F{{F}}
\def\FF{{\cal F}}
\parn
\medn
{\bf Abstract.} \quad
The $q$-vertex operators of Frenkel and Reshetikhin are studied by means of a
$q$-deformation of the Wakimoto module for the quantum affine algebra
$U_q(\widehat{\sl}_2)$ at an arbitrary level $k\ne 0,-2$.
A Fock module version of the $q$-deformed primary field of spin $j$ is
introduced, as well as the screening operators which (anti-)commute  with the
action of $U_q(\widehat{\sl}_2)$ up to a total difference of a field.
A proof of the intertwining property is given for the $q$-vertex operators
corresponding to the primary fields of spin $j\notin {1 \over2}\Z_{\geq0}$,
which is enough to treat a general case.
A sample calculation of the correlation function is also given.
\vbn
\beginsection{1. Introduction}\par
In a recent paper [FR], Frenkel and Reshetikhin constructed a certain
$q$-deformation of the Wess-Zumino-Witten (WZW) model on the sphere in the
operator formalism based on the representation theory of the quantum affine
algebras.
They defined the notion of $q$-deformed chiral vertex operators as certain
intertwining operators, which give an analogue of the primary fields.
In principle, the intertwining property characterizes them, however, it is not
easy to find an explicit expression of them.
\par
For $\widehat{\sl}_2$ WZW model, the following realization is known (cf.
[FLMSS]).
The standard $\sl_2$ currents $J^\pm(z)$, $J^0(z)$, screening operators $S(z)$,
$S^+(z)$ and the primary fields $\phi_{j,m}(z)$ of spin $j$ are explicitly
written as
$$\matrix{
J^\pm(z)=\,:{1 \over{\sqrt{2}}}\left[\sqrt{k +2}\partial\varphi_1(z)\pm
i\sqrt{k}\partial\varphi_2(z)\right]e^{\pm\sqrt{{{2 \over
k}}}[i\varphi_2(z)-\varphi_0(z)]}:,\cr\cr
J^0(z)=-\sqrt{\ds{k \over2}}\partial\varphi_0(z),\cr\cr
S(z)=\,:-{1 \over{\sqrt{2}}}\left[\sqrt{k +2}\partial\varphi_1(z)
+i\sqrt{k}\partial\varphi_2(z)\right]e^{-\sqrt{{{2 \over {k
+2}}}}\varphi_1(z)}:,\cr\cr
S^+(z)=:e^{\left[\sqrt{{{k +2} \over {2}}}\varphi_1(z) +\sqrt{{k \over
2}}i\varphi_2(z)\right]}:\cr\cr
\phi_{j,m}(z)=:e^{\left[j\sqrt{{2 \over{k +2}}}\varphi_1(z) +m\sqrt{{2 \over
k}}(i\varphi_2(z)-\varphi_0(z))\right]}:\cr\cr
}\no(1.1)$$
where $\varphi_i(z)$ are independent bosonic fields normalized as
$\varphi_i(z)\varphi_i(w)\sim \log(z-w)$.
Then the operator product of $S(z)$ or $S^+(z)$ with the current is a total
derivative of a field, which is anihilated by integration on a closed cycle.
Note that by fixing the picture of the Fock module [FMS] and restricting it to
$\ker\ Q^+$, where $Q^+=\oint S^+(z)dz$, one obtains the Wakimoto module [W].
Among these Fock modules the primary fields act and a combination of them with
the screening charge, which is an integration of the composite operator
$S(t_1)\cdots S(t_r)$ on a certain cycle, gives rise to the chiral vertex
operator of Tsuchiya and Kanie [TK], see also [BF].
\par
The aim of the present article is to construct a $q$-deformation of this
realization following the line of a previous work [M3]: the operators are
deformed by changing the normalization of the modes of $\varphi_i(z)$ and by
replacing the differential $\partial\varphi_i(z)$ with a certain difference.
In this paper we show that $S(z)$, $S^+(z)$ and $\phi_{j,m}(z)$ are also
deformed in the same spirit.
Then the operator $Q^+=\oint S^+(z) dz$ (anti)-commute with the currents and
the primary fields.
We also fix the picture and restrict the Fock module to $\ker\ Q^+$.
Thus we obtain a Fock module version of the $q$-deformed chiral vertex
operators of Frenkel and Reshetikhin.
We show that the operator product of $S(z)$ with the current is a total
$p$-difference of a field, where $p=q^{2(k +2)}$.
This total difference is eliminated by the Jackson integral, a $p$-deformation
of integration, thus we obtain a screening charge.
Therefore, putting aside a problem of choice of a $q$-cycle of the Jackson
integral, we are able to obtain an explicit expression of the correlation
function,
which gives a Jackson integral solution to the quantum ($q$-deformed)
Knizhnik-Zamolodchikov equation.
\par
The paper is organized as follows.
In sect.2, we summarize some notions from representation theory of the quantum
affine algebra $U_q(\widehat{\sl}_2)$, and fix the notations in $q$-analysis.
In sect.3 we review the result of [M3] in a modified form.
Sect.4 is devoted to a construction of a deformation of the screening operators
$S(z)$ and $S^+(z)$.
We also discuss the restriction of the Fock spaces.
In sect.5, the primary fields of spin $j$ are introduced and the intertwining
property of the corresponding $q$-vertex operator is proved except for
$2j\in\Z_{\geq 0}$.
In sect.6 we discuss a property of the correlation function, and give a sample
calculation of it.
A brief conclusion will be given in sect. 7.
\par
Recently Kato et al.\ [KQS] also discussed a free field realization of
$q$-vertex operators using Shiraishi's representation of $U_q(\widehat{\sl}_2)$
[S].
A comment on [KQS] will also be given in sect. 7.

\beginsection{2. Preliminaries}\par
Let $\C$ denote the field of complex numbers and $\C^*=\C-\{0\}$ its
multiplicative group.
Let $q$ be a complex number transcendental over ${\bf Q}$, the field of
rational numbers.
In the sequel, if necessary, we choose an appropriate branch of the complex
power of the form $q^{\mu}$, $\mu\in \C$.
\par
The quantum affine algebra $\U=U_q(\widehat{\sl}_2)$ is the associative algebra
generated over $\C$ by the letters $e_0,e_1,f_0,f_1,q^{\pm h_0},q^{\pm
h_1},q^{\pm d}$ satisfying the following defining relations:
$$\eqalign{
q^{h_0}q^{h_1}=q^{h_1}q^{h_0},\ q^dq^{h_i}&=q^{h_i}q^d,\
q^{h_0}q^{-h_0}=q^{h_1}q^{-h_1}=q^dq^{-d}=1,\cr
q^{h_i}e_iq^{-h_i}=&q^{2}e_i,\ q^{h_i}e_jq^{-h_i}=q^{-2}e_j\ (i\ne j),\cr
q^{h_i}f_iq^{-h_i}=&q^{-2}f_i,\ q^{h_i}f_jq^{-h_i}=q^{2}f_j\ (i\ne j),\cr
q^de_iq^{-d}=&q^{\delta_{i,0}}e_i,\ q^df_iq^{-d}=q^{-\delta_{i,0}}f_i,\cr
&[e_i,\ f_j]=\delta_{ij}{{q^{h_i}-q^{-h_i}} \over{q-q^{-1}}},\cr
e_i^3e_j-(q^2 +&1 +q^{-2})e_i^2e_je_i\cr
  +(q^2 +&1 +q^{-2})e_ie_je_i^2-e_je_i^3=0\ (i\ne j),\cr
f_i^3f_j-(q^2 +&1 +q^{-2})f_i^2f_jf_i\cr
  +(q^2 +&1 +q^{-2})f_if_jf_i^2-f_jf_i^3=0\ (i\ne j).\cr}\no(2.1)$$
where we have used the standard notation:
$$[m]={{q^m-q^{-m}} \over{q-q^{-1}}}.\no(2.2)$$
\par
The subalgebra generated by $e_0,e_1,f_0,f_1,q^{\pm h_0},q^{\pm h_1}$ is
denoted by $\U'$.
This algebra admits another realization in terms of generators and relations,
called the Drinfeld realization [D2], which will play a crucial in our
construction [M3].
However, since apparently it is not necessary in this paper, we ommit it.
\par
The algebra $\U$ becomes a bialgebra with the comultiplication defined by
$$\eqalign{
&\Delta(e_i)=e_i\otimes1 +q^{h_i}\otimes e_i,\cr
&\Delta(f_i)=f_i\otimes q^{-h_i} +1\otimes f_i,\cr
&\Delta(q^{h_i})=q^{h_i}\otimes q^{h_i},\cr
&\Delta(q^d)=q^d\otimes q^d.\cr
}\no(2.3)$$
Thus we can consider the tensor product of two representations.
Note that this choice of comultiplication coinsides with that of [JMMN] or
[DFJMN] and differs from that of [FR].
\par
Let us introduce two kinds of representation of $\U$ for later use.
The first kind is the highest weight representations (cf. [L]) and the other is
the evaluation representations (cf. [J]).
The former are not necessarily integrable, and the latter are not necessarily
the affinization of finite dimensional modules.
\par
A $\U$-module $M$ is said to be a highest weight module of spin $l$ at level
$k$ if there exists a vector $v_l\in M$ such that
$$M=\U v,\ e_1v=e_0v=0,\ q^{h_1}v=q^{2l}v,\ q^{h_0}v=q^{k-2l}v\and q^d v=v.
\no(2.4)$$
The vector $v$ is called the highest weight vector.
Then the highest weight of $M$ is $2l\Lambda_1 +(k-2l)\Lambda_0$, where
$\Lambda_i$, $i=0,1$, are the usual fundamental weights.
For instance the Verma module of $\U$ is a highest weight module similarly
defined as in the case of a Kac-Moody Lie algebra (cf. [L]).
It is irreducible unless the highest weight satisfies a special condition.
\par
The evaluation representation concerned in this paper is the following type:
the vector space $V_j(z)=\left(\oplus_{l=0}^\infty \C v_{j,j-l}\right)\otimes
\C[z,z^{-1}]$ equipped with the $\U$-module structure defined by
$$\eqalign{
e_1v_{j,m}\otimes z^n&=[j +m]v_{j,m-1}\otimes z^n,\cr
e_0v_{j,m}\otimes z^n&=[j-m]v_{j,m +1}\otimes z^{n +1},\cr
f_1v_{j,m}\otimes z^n&=[j-m]v_{j,m +1}\otimes z^n,\cr
f_0v_{j,m}\otimes z^n&=[j +m]v_{j,m-1}\otimes z^{n-1},\cr
q^{h_1}v_{j,m}\otimes z^n&=q^{j-2m}v_{j,m}\otimes z^n,\cr
q^{h_0}v_{j,m}\otimes z^n&=q^{2m-j}v_{j,m}\otimes z^n,\cr
q^dv_{j,m}\otimes z^n&=q^{n}v_{j,m}\otimes z^n.\cr
}\no(2.5)$$
The space $V_j(z)$ is also endowed with the natural $\C[z,z^{-1}]$-module
structure.
When $2j$ is a non-negative integer, then $\left(\oplus_{l=0}^{2j} \C
v_{j,j-l}\right)\otimes \C[z,z^{-1}]$ is the affinization of a finite
dimensional $\U$-module [J1].
\par
Now let us turn to preliminaries from $q$-analysis.
We define the difference operator ${{\partial_{p}} \over{\partial_{p}z}}$ for a
scalar $p\in\C^*$ by:
$${{\partial_{p}} \over{\partial_{p}z}}f(z)={{f(p^{1/2}z)-f(p^{-1/2}z)} \over
{(p^{1/2}-p^{-1/2})z}}\no(2.6)$$
for a function $f(z)$ on $\C^*$.
A function of the form ${{\partial_{p}} \over{\partial_{p}z}}f(z)$ is called a
{\it total (p-)difference} of a function $f(z)$.
\par
To eliminate a total difference, we need some analogue of integration.
For instance, the Jackson integral, defined by
$$\int_0^{s\infty}f(z){{d_pt} \over
z}=(1-p)\sum_{p=-\infty}^{\infty}f(sp^n)\no(2.7)$$
for a scalar $s\in\C^*$, satisfies
$$\int_0^{s\infty} \left\{{{\partial_p} \over
{\partial_pz}}f(z)\right\}{{d_pz}}=0\no(2.8)$$
if it is convergent.
Another example is to take the residue at zero:
$${1 \over{2\pi\sqrt{-1}}}\oint_{z=0} f(z){{dz}}=\left[{\hbox{ residue of
$f(z)$ at
$z=0$}}\right],\no(2.9)$$
for a meromorphic function.
Then we have
$${1 \over{2\pi\sqrt{-1}}}\oint_{z=0} {{\partial_p} \over {\partial_p
z}}f(z){{dz}}=0\no(2.10)$$
\par
We finally note that in this paper we shall freely use the technique of {\it
operator product expansion}.
Let $A_n:\F_1\rightarrow \F_2$, $n\in\Z$, be linear maps of vector spaces
$F_1$, $F_2$.
Then the power series $A(z)=\sum_{n\in\Z}z^{-n -\Delta} A_n$, where $\Delta$ is
a complex number, is simply called an operator or a field of dimension
$\Delta$.
We will write by abuse of notation as $A(z):F_1\rightarrow F_2$.
Now let $A(z):F_2\rightarrow F_3$ and $B(z):F_1\rightarrow F_2$ be operators.
Then the composition $A(z)B(w):F_1\rightarrow F_3$ is defined as formal power
series.
When it can be analytically continued outside some locus, the notation
$A(z)B(w)\sim C(z,w)$ means that $A(z)B(w)-C(z,w)$ is holomorphic on
$\C^*\times\C^*$ in appropriate sense.
Note that the singular locus of $C(z,w)$ is not necessarily $z=w$ in our
construction, unlike the $q=1$ case.
The reader who is not familiar with this technique should be referred to [TK]
for a mathematical survey.

\beginsection{3. Free field representation of $U_q(\widehat{\sl}_2)$}\par
Let $\{\a_n,\bar \a_n,\b_n\,|\,n\in\Z\}$ be a set of operators satisfying the
following commutation relations:
$$\eqalign{
&[\a_m,\,\a_{-m}]={{[2m][km]} \over {m}},\cr
&[\bar \a_m,\,\bar \a_{-m}]=-{{[2m][km]} \over {m}},\cr
&[\b_m,\,\b_{-m}]={{[2m][(k +2)m]} \over {m}}.\cr
}\no(3.1)$$
Suppose that the other commutators are zero.
By renormalizing these operators, we have the usual modes of three bosonic
fields.
In other words they form the direct sum of three Heisenberg algebras with
infinite generators.
\par
We define
$$\eqalign{
N_ +&=\C[\a_m,\bar \a_m,\b_m]_{m>0},\cr
N_-&=\C[\a_m,\bar \a_m,\b_m]_{m<0}.\cr}\no(3.2)$$
The left Fock module $F_{l,m_1,m_2}$ is uniquely characterized by the following
properties:
there exists a vector $|l,m_1,m_2\>$ in $\F_{l,m_1,m_2}$ such that
$$\eqalign{
&\b_0|l,m_1,m_2\>=2l\,|l,m_1,m_2\>, \cr
&\a_0|l,m_1,m_2\>=2m_1|l,m_1,m_2\>, \cr
&\bar \a_0|l,m_1,m_2\>=-2m_2|l,m_1,m_2\>,\cr
&N_ +|l,m_1,m_2\>=0, \and\cr
&N_-|l,m_1,m_2\>{\hbox{ is a free $N_-$-module of rank 1.}}\cr}\no(3.3.L)$$
The right Fock module $F_{l,m_1,m_2}^{\dagger}$ is similarly characterized by
$$\eqalign{
&\<l,m_1,m_2|\b_0=2l\,\<l,m_1,m_2|, \cr
&\<l,m_1,m_2|\a_0=2m_1\<l,m_1,m_2|, \cr
&\<l,m_1,m_2|\bar \a_0=-2m_2\<l,m_1,m_2|,\cr
&\<l,m_1,m_2|N_-=0, \and\cr
&\<l,m_1,m_2|N_ +{\hbox{ is a free $N_ +$-module of rank 1.}}\cr}\no(3.3.R)$$
In the sequel we will be mainly concerned with the left Fock modules.
Each statement will have a right module counterpart.
\par
For each triple of complex numbers $r$, $s_1$ and $s_2$, we define the operator
$e^{2r\b +2s_1\a +2s_2\bar \a}$ by the mapping $|l,m_1,m_2\>\mapsto |l +r,m_1
+s_1,m_2 +s_2\>$ such that it commutes with the action of $N_\pm$.
The normal ordering $:\ :$ is defined according to $\a<\a_0$, $\bar \a<\bar
\a_0$, $\b <\b_0$ and $N_-<N_ +$.
\par
Consider the operators $X^\pm(z)\,:\,\F_{l,m_1,m_2}\rightarrow
\F_{l,m_1\pm1,m_2\pm1}$ defined by
$$\eqalign{
X^{+}(z)&={1 \over{(q-q^{-1})z}}:Y^+(z)\biggl\{Z_ +(q^{-{{k +2} \over2}}z)W_
+(q^{-{k \over2}}z)-W_-(q^{{k \over2}}z)Z_-(q^{{{k +2} \over2}}z)\biggr\}:,\cr
X^{-}(z)&={-1 \over{(q-q^{-1})z}}:Y^-(z)\biggl\{Z_ +(q^{{{k +2} \over2}}z)W_
+(q^{{k \over2}}z)^{-1}-W_-(q^{-{k \over2}}z)^{-1}Z_-(q^{-{{k +2}
\over2}}z)\biggr\}:,\cr}\no(3.4)$$
where
$$\eqalign{
Y^{+}(z)
 &=\exp\left\{\sum_{m=1}^\infty q^{-{{km} \over 2}}{{z^m}
\over{[km]}}\bigl(\a_{-m} +\bar \a_{-m}\bigr)\right\}\cr
 &\qquad e^{2(\a +\bar \a)}z^{{1 \over k}(\a_0 +\bar
\a_0)}\exp\left\{-\sum_{m=1}^\infty q^{-{{km} \over2}}{{z^{-m}}
\over{[km]}}\bigl(\a_m +\bar \a_m\bigr)\right\},\cr
Y^{-}(z)
 &=\exp\left\{-\sum_{m=1}^\infty q^{{{km} \over 2}}{{z^m}
\over{[km]}}\bigl(\a_{-m} +\bar \a_{-m}\bigr)\right\}\cr
 &\qquad e^{-2(\a +\bar \a)}z^{-{1 \over k}(\a_0 +\bar
\a_0)}\exp\left\{\sum_{m=1}^\infty q^{{{km} \over2}}{{z^{-m}}
\over{[km]}}\bigl(\a_m +\bar \a_m\bigr)\right\},\cr}\no(3.5)$$
$$\eqalign{
Z_ +(z)&=\exp\left\{-(q-q^{-1})\sum_{m=1}^\infty z^{-m}{{[m]} \over{[2m]}}\bar
\a_{m}\right\}q^{-{1 \over 2}\bar \a_{0}},\cr
Z_-(z)&=\exp\left\{(q-q^{-1})\sum_{m=1}^\infty z^{m}{{[m]} \over{[2m]}}\bar
\a_{-m}\right\}q^{{1 \over 2}\bar \a_{0}},\cr}\no(3.6)$$
$$\eqalign{
W_{+}(z)&=\exp\left\{-(q-q^{-1})\sum_{m=1}^\infty z^{-m}{{[m]}
\over{[2m]}}\b_{m}\right\}q^{-{1 \over 2}\b_{0}},\cr
W_{-}(z)&=\exp\left\{(q-q^{-1})\sum_{m=1}^\infty z^{m}{{[m]}
\over{[2m]}}\b_{-m}\right\}q^{{1 \over 2}\b_{0}}.\cr}\no(3.7)$$
\medn
Now $\a_0 +\bar \a_0$ acts on $\F_{l,m,m}$ by zero.
Therefore the expansion of the form
$$X^\pm(z)=\sum_{m\in\Z}x^\pm_m z^{-m}\no(3.10)$$
makes sense on $\F_{l,m,m}$.
Moreover let us introduce the operator $p^\d$ for a scalar $p\in\C^{*}$
characterized by
$$p^\d\a_mp^{-\d}=p^m\a_m,\ p^\d\bar \a_mp^{-\d}=p^m\bar \a_m,\
p^\d\b_mp^{-\d}=p^m\b_m,\no(3.11)$$
and
$$p^\d|l,m_1,m_2\>=1 {\hbox{ for any $l,m_1,m_2$.}}\no(3.12)$$
Then we have
\medn
{\bf Proposition 3.1 }
There exists a representation $\pi_l\,:\,\U\rightarrow \End(\oplus_{m=l +\Z}
\F_{l,m,m})$ such that
$$\matrix{
&\pi_l(e_1)=x_0^+,
\hfill&\pi_l(f_1)=x_0^-,
\hfill&\pi_l(q^{h_1})=q^{\a_0},\hfill\cr\cr
&\pi_l(e_0)=x_1^-q^{-\a_0},
\hfill&\pi_l(f_0)=q^{\a_0}x_{-1}^+,
\hfill&\pi_l(q^{h_0})=q^{k}q^{-\a_0},
\hfill&\pi_l(q^d)=q^\d.\hfill\cr\cr}$$
{\it Proof. } Straightforward, see [M3].\QED
\medn
{\it Remark 1.} To make (3.4) be single valued, ${1 \over k}(\a_0 +\bar \a_0)$
is necessarily an integer.
The constraint $\a_0 +\bar \a_0=0$ means that we have fixed a {\it picture} of
Fock representation, see [FMS] and [FLMSS].
\parn
{\it Remark 2.} Let $L_0$ be the operator defined by
$$\eqalign{
L_0&=\sum_{m>0}{{m^2} \over{[2m][km]}}(\a_{-m}\a_{m}-\bar\a_{-m}\bar\a_{m}) +{1
\over{4k}}(\a_0^2-\bar\a_0^2)\cr
&\quad +\sum_{m>0}{{m^2} \over{[2m][(k +2)m]}}\b_{-m}\b_{m} +{1 \over{4(k
+2)}}(\b_0^2 +2\b_0)\cr
}\no(3.13)$$
Then we have $\delta +L_0={{l(l +1)} \over {k +2}}$ on $\F_{l,m_1,m_2}$.
When $q=1$, this is the usual $L_0$ of the energy momentum tensor
$T(z)=\sum_{m\in\Z}z^{-m-2}L_m$ (cf.\ [FLMSS]).
\med
In the sequel we will drop $\pi_l$ when we write the action of$\U$.
\medn
{\bf Proposition 3.2 }
The vector $|l,l,l\>$ for any $l\in \C$ satisies $e_1|l,l,l\>=e_0|l,l,l\>=0$.
\medn
{\it Proof. } Straightforward.\QED

\beginsection{4. Screening operators and structure of Fock spaces}\par
In this section we define an analogue of screening operators.
We set
$$H(z)=\sum_{m\in\Z}z^{-m-1}\a_m\no(4.1)$$
for convenience.
\par
Consider the operator $S^+(z)\,:\,\F_{l,m_1,m_2}\rightarrow \F_{l +{{k +2}
\over2},m_1,m_2 +{k \over2}}$ defined by:
$$\eqalign{
S^+(z)
 &=\exp\left\{\sum_{m=1}^\infty z^m{1 \over{[2m]}}\left(q^{{{k} \over
2}m}\b_{-m} +q^{{{k +2} \over 2}m}\bar \a_{-m}\right)\right\}\cr
 &\qquad e^{(k +2)\b +k\bar \a} z^{{1 \over2}(\b_{0} +\bar
\a_0)}\exp\left\{-\sum_{m=1}^\infty z^{-m}{1 \over{[2m]}}\left(q^{{{k} \over
2}m}\b_{m} +q^{{{k +2} \over 2}m}\bar \a_m\right)\right\}\cr}\no(4.2)$$
\medn
{\bf Lemma 4.1 } We have the following relations:
$$\eqalign{
&X^+(z)S^+(w)=-S^+(w)X^+(z)\sim {{\partial_q} \over{\partial_qw}} \left\{{1
\over {z-w}}Y^+(z) S_1^+(z)\right\},\cr
&X^-(z)S^+(w)=-S^+(w)X^-(z)\sim0,\cr
&H(z)S^+(w)=S^+(w)H^-(z)\sim0,\cr
&q^dS^+(w)q^{-d}=q^{m_2-l}S^+(q^{-1}w),\cr
}$$
where
$$\eqalign{
S_1^+(z)
 &=\exp\left\{\sum_{m=1}^\infty z^m{1 \over{[2m]}}\left(q^{{{k +2} \over
2}m}\b_{-m} +q^{{{k +4} \over 2}m}\bar \a_{-m}\right)\right\}\cr
 &\qquad e^{(k +2)\b +k\bar \a} z^{{1 \over2}(\b_{0} +\bar
\a_0)}\exp\left\{-\sum_{m=1}^\infty z^{-m}{1 \over{[2m]}}\left(q^{{{k +2} \over
2}m}\b_{m} +q^{{{k +4} \over 2}m}\bar \a_m\right)\right\}.\cr}$$
\medn
{\it Proof. } Straightforward.
\QED
\med
Note that $S^+(z)$ is single valued on $\F_{l,m_1,m_2}$ provided $l-m_2$ is an
integer.
Let $Q^+$ be defined by:
$$Q^+={1 \over{2\pi \sqrt{-1}}}\oint S^+(z){{dz}}.\no(4.3)$$
Then we immediately see
\medn
{\bf Proposition 4.2 } The operator $Q^+$ commutes with the action of $\U'$ up
to sign and with $q^d$ up to a scalar multiple.
\med
Now we are in a position to consider a restriction of the Fock spaces.
We define
$$\F_l=\oplus_{m=l +\Z}\ker\,(Q^+:\F_{l,m,m}\rightarrow\F_{l +{{k +2}
\over2},m,m +{k \over2}}).\no(4.4)$$
Because of Proposition 4.2, $\U$ acts on $\F_l$.
Moreover the following holds:
\medn
{\bf Proposition 4.3 }
The character of $\F_l$ is same as the Verma module of spin $l$ at level $k$.
\medn
{\it Proof. }
Let us first make the following observation:
putting
$$S^+(z)=\,:\exp \chi(z):$$
where
$$\eqalign{
\chi(z)
&=\sum_{m=1}^\infty z^m\left\{{q^{{k \over2}m} \over{[2m]}}\b_m +{q^{{{k +2}
\over2}m} \over{[2m]}}\bar \a_m\right\} +(k +2)\b +k\bar\a\cr
& +{1 \over 2}\log(z)(\b_0 +\bar \a_0)
-\sum_{m=1}^\infty z^{-m}\left\{{q^{{k \over2}m} \over{[2m]}}\b_m +{q^{{{k +2}
\over2}m} \over{[2m]}}\bar \a_m\right\},\cr}$$
we have $\chi(z)\chi(w)\sim \log(z-w)$.
Therefore we may understand $Q^+$ as the zero mode $\eta_0$ of the fermionic
ghost system $(\eta,\xi)$ of dimension (1,0):
$$\eta(z)=\sum_{m\in\Z}z^{-m-1}\eta_m=:e^{\chi(z)}:,\quad
\xi(z)=\sum_{m\in\Z}z^{-m}\xi_m=:e^{-\chi(z)}:.$$
Since we have $\eta_0^2=0$ and $\eta_0\xi_0 +\xi_0\eta_0=1$, we obtain the
following exact sequence:
$$\matrix{
&Q^+&&Q^+&&Q^+\cr
\cdots&\rightarrow &\oplus_{m}\F_{l,m,m}&\rightarrow &\oplus_m\F_{l +{{k +2}
\over 2},m,m +{k \over 2}}&\rightarrow&\cdots\cr}$$
By using this sequence, we can compute the character of $\F_l$, which is shown
to be same as the Verma module with spin $l$ at level $k$.
\QED
\medn
In particular we have
\medn
{\bf Corollary 4.4 } Let $l$ be a general complex number such that the Verma
module with spin $l$ at level $k$ is irreducible.
Then it is isomorphic to $\F_l$ with the highest weight vector $|l,l,l\>$.
\medn
{\it Note. }
A detail of the discussion of the character is communiacted to the author by A.
Tsuchiya.
When $q=1$, the result is also found in some physics literatures, see [FMS],
[FLMSS].
In this limit, $\F_l$ is isomorphic to the Wakimoto module of $\widehat{\sl}_2$
(cf. [FLMSS]).
This situation allows us to claim that our representation is a $q$-deformation
of the Wakimoto module.
\medn
{\it Remark. }
There exists another screening operator $S^-(z)$ defined by
$$\eqalign{
S^-(z)
 &=\exp\left\{-\sum_{m=1}^\infty z^m{1 \over{[2m]}}\left(q^{-{{k} \over
2}m}\b_{-m}-q^{-{{k +2} \over 2}m}\bar \a_{-m}\right)\right\}\cr
 &\qquad e^{-(k +2)\b +k\bar \a} z^{{1 \over2}(-\b_{0} +\bar
\a_0)}\exp\left\{\sum_{m=1}^\infty z^{-m}{1 \over{[2m]}}\left(q^{-{{k} \over
2}m}\b_{m}-q^{-{{k +2} \over 2}m}\bar \a_m\right)\right\}.\cr}$$
\med
Now until the end of this section, we consider the other screening operator.
Consider the operator $S(z)\,:\,\F_{l,m_1,m_2}\rightarrow \F_{l-1,m_1,m_2}$
defined as follow:
$$\eqalign{S(z)&={{-1} \over{(q-q^{-1})z}}:U(z)\biggl\{Z_ +(q^{-{{k +2}
\over2}}z)^{-1}W_ +(q^{-{k \over2}}z)^{-1}-W_-(q^{{k \over2}}z)^{-1}Z_-(q^{{{k
+2} \over2}}z)^{-1}\biggr\}:,\cr}\no(4.5)$$
where $Z_\pm(z)$ and $W_\pm(z)$ are defined by (3.6) and (3.7) respectively,
and
$$\eqalign{
U(z)
 &=\exp\left\{-\sum_{m=1}^\infty z^m{q^{-{{k +2} \over 2}m} \over{[(k
+2)m]}}\b_{-m}\right\}\cr
 &\qqquad e^{-2\b} z^{-{{1} \over {k +2}}\b_{0}}\exp\left\{\sum_{m=1}^\infty
z^{-m}{q^{-{{k +2} \over2}m} \over{[(k +2)m]}}\b_{m}\right\}.\cr}\no(4.6)$$
\medn
{\bf Lemma 4.5 } We have the following relations
$$\eqalignno{
&S^+(z)S(w)=S(w)S^+(z)\sim0,\cr
&X^+(z)S(w)=S(w)X^+(z)\sim0\cr
&X^-(z)S(w)=S(w)X^-(z)\sim [k +2]{{\partial_{p}} \over{\partial_{p}w}}\left\{{1
\over{z-w}}Y^-(z)U_1(z)\right\},\cr
&H(z)S(w)=S(w)H(z)\sim0,\cr
&p^{d}S(w)p^{-d}=p^{-{{2l} \over{k +2}} +1}S(p^{-1}w),\cr
}$$
where $p=q^{2(k +2)}$, $Y^-(z)$ is defined by (3.5), and
$$\eqalign{
U_1(z)
 &=\exp\left\{-\sum_{m=1}^\infty z^m{q^{{{k +2} \over 2}m} \over{[(k
+2)m]}}\b_{-m}\right\}\cr
 &\qqquad e^{-2\b} z^{-{1 \over {k +2}}\b_{0}}\exp\left\{\sum_{m=1}^\infty
z^{-m}{q^{{{k +2} \over2}m} \over{[(k +2)m]}}\b_{m}\right\}.\cr}$$
\parn
{\it Proof. } Straightforward.\QED
\medn
Thus $S(z)$ defines an operator
$$S(z):\F_l\rightarrow \F_{l-1},\no(4.7)$$
and we have
\medn
{\bf Proposition 4.6 } The operator $S(z)$ commutes with the action of $\U'$ up
to total $p$-difference of a field.
\medn
{\it Remark. }
Let $L_0$ be the operator defined by (3.13).
Then we have
$$p^{L_0}S(w)p^{-L_0}=pS(pw)=S(w) +(p-1){{\partial_p}
\over{\partial_pw}}\left\{wS(p^{{1 \over 2}}w)\right\}.$$
Therefore the operator $S(w)$ commutes with $p^{L_0}$ modulo a total
$p$-difference.
When $q=1$, it corresponds to $[L_0, S(w)]={{\partial } \over{\partial
w}}\left\{wS(w)\right\}$, which comes from the operator product of $S(w)$ with
the energy momentum tensor $T(z)=\sum_{m\in\Z}z^{-m-2}L_m$.

\beginsection{5. Primary fields and $q$-vertex operators}\par
In this section we construct a Fock module version of $q$-vertex operators.
As for the formulation of intertwining operators, we follow [J2] (cf. [JMMN],
[IIJMNT]).
For any complex number $j$, let $\phi_j(z)\,:\,\F_{r,m_1,m_2}\rightarrow \F_{r
+j,m_1 +j,m_2 +j}$ be the operator defined by
$$\eqalign{
\phi_j(z)=
&\exp\left\{\sum_{m=1}^\infty (q^{k +2}z)^m{{q^{{{km} \over 2}}[2jm]}
\over{[km][2m]}}\bigl(\a_{-m} +\bar \a_{-m}\bigr)\right\}\cr
 &e^{2j(\a +\bar \a)}z^{{{j} \over{k +2}}(\a_{0} +\bar
\a_{0})}\exp\left\{-\sum_{m=1}^\infty (q^{k +2}z)^{-m}{{q^{{{km} \over2}}[2jm]}
\over{[km][2m]}}\bigl(\a_{m} +\bar \a_{m}\bigr)\right\}\cr
&\exp\left\{\sum_{m=1}^\infty (q^{k +2}z)^m{{q^{{{k +2} \over 2}m}[2jm]}
\over{[2m][(k +2)m]}}\b_{-m}\right\}\cr
 &\qqquad e^{2j\b} z^{{{j} \over {k +2}}\b_{0}}\exp\left\{-\sum_{m=1}^\infty
(q^{k +2}z)^{-m}{{q^{{{k +2} \over2}m}[2jm]} \over{[2m][(k
+2)m]}}\b_{m}\right\}.\cr
}\no(5.1)$$
The primary fields of spin $j$ are defined inductively by
$$\eqalign{
&\phi_{j,m-1}(z)={1 \over{[j-m
+1]}}\left\{\phi_{j,m}(z)x^-_0-q^{2m}x^-_0\phi_{j,m}(z)\right\},\cr
&\phi_{j,j}(z)=\phi_j(z).\cr}\no(5.2)$$
We understand $\phi_{j,m}(z)=0$ if $j-m\notin \Z_{\geq 0}$.
\medn
{\bf Proposition 5.1 } The primary field $\phi_{j,m}(z)$ commutes with the
action of $Q^+$ up to sign.
\medn
{\it Proof. }
The statement follows from Corollary 4.3 and
$S^+(z)\phi_j(w)=\phi_j(w)S^+(z)\sim 0$, which is proved by a straightforward
calculation.
\QED
\medn
Therefore the image of $\F_l$ by each component of $\phi_{j,m}(z)$ is contained
in $\F_{l +j}$ for any $l\in \C$.
Thus we have obtained an operator
$$\phi_{j,m}(z)\,:\,\F_{l}\longrightarrow \F_{l +j}.\no(5.3)$$
\med
Now let us consider the intertwining property of the primary fields.
\medn
{\bf Lemma 5.2 }
By analytic continuation, we have the following:
$$\leqalignno{
&X^+(w)\phi_j(z)=\phi_j(z)X^+(w)\sim 0,&\ \ (1)\cr
&(q^{k +2}z-q^{2j}w)\phi_j(z)X^-(w)=(q^{2j +k +2}z-w)X^-(w)\phi_j(z)\sim 0.&\ \
(2)\cr}$$
\medn
{\it Proof. }
Straightforward.\QED
\medn
{\bf Proposition 5.3 }
We have the following relations:
$$\leqalignno{
&\phi_{j,m}(z)e_1=e_1\phi_{j,m}(z) +[j +m +1]q^{h_1}\phi_{j,m +1}(z),&\ \
(1)\cr
&\phi_{j,m}(z)e_0=e_0\phi_{j,m}(z) +z[j-m +1]q^{h_0}\phi_{j,m-1}(z),&\ \ (2)\cr
&\phi_{j,m}(z)f_1=q^{2m}f_1\phi_{j,m}(z) +[j-m +1]\phi_{j,m-1}(z),&\ \ (3)\cr
&\phi_{j,m}(z)f_0=q^{-2m}f_0\phi_{j,m}(z) +z^{-1}[j +m +1]\phi_{j,m +1}(z),&\ \
(4)\cr
&\phi_{j,m}(z)q^{h_1}=q^{-2m}q^{h_1}\phi_{j,m}(z),&\ \ (5)\cr
&\phi_{j,m}(z)q^{h_0}=q^{2m}q^{h_0}\phi_{j,m}(z),&\ \ (6)\cr
&\phi_{j,m}(z)q^d=q^{d-{{2jl} \over{k +2}}}\phi_{j,m}(qz).&\ \ (7)\cr
}$$
\medn
{\it Proof. }
The relation (3) is obvious by the defintion, and (5) (6) and (7) are directly
proved.
The relations (1), (2) and (4) for $m=j$ are checked by Lemma 5.3.
Then the relation (2) for $m<j$ is proved by induction on $m$.
Finally (1) and (4) are derived from (2) and (3).
\QED
\medn
Let us put
$$\Delta_l={{l(l +1)} \over{k +2}}.\no(5.4)$$
As a consequence of Proposition 5.3, we have the following theorem.
\medn
{\bf Theorem 5.4 }
Suppose that $2j\notin \Z_{\geq 0}$.
Then the operator
$$\tilde \Phi_{l,j}^{l +j}(z)=z^{\Delta_l +\Delta_j-\Delta_{l
+j}}\sum_{m=0}^\infty\phi_{j,j-m}(z)\otimes v_{j,j-m},$$
gives rise to an intertwining operator $\F_{l}\rightarrow \F_{l +j}\otimes
V_j(z)$ of $\U$-modules.
\medn
{\it Remark 1. }
This theorem asserts that $\tilde\Phi_{j,l}^{l +j}(z)$ is a Fock module version
of the $q$-deformed chiral vertex operators of Frenkel and Reshetikhin.
Note that to prove the intertwining property when $2j\in\Z_{\geq 0}$ one must
verify $\phi_{j,-j}(z)f_1=q^{-2j}f_1\phi_{j,-j}(z)$ and
$\phi_{j,-j}(z)e_0=e_0\phi_{j,-j}(z)$.
\medn
{\it Remark 2. }
We formally set
$$\Phi_{l,j}^{l +j-r}(z)=z^{\Delta_l +\Delta_j-\Delta_{l
+j-r}}\sum_{m=0}^\infty
\int_0^{s_1\infty}S(t_1){d_pt_1}\cdots\int_0^{s_{r}\infty}S(t_{r}){d_pt_r}
\phi_{j,j-m}^{(r)}(z)\otimes v_{j,j-m}$$
for an appropriate choice of scalars $s_1,\cdots,s_r$.
Then it would give an intertwining operator
$$\tilde \Phi_{l,j}^{l +j-r}(z)\,:\,\F_{l}\rightarrow \F_{l +j-r}\otimes
V_j(z)$$
for an arbitrary $l\in\C^*$.
To understand it mathematically, we need to study a choice of the $q$-cycle of
the Jackson integral rigorously in the operator formalism.

\beginsection{6. Correlation functions}\par
Let $j_0,\cdots,j_N,j_\infty$ be a set of complex numbers satisfying
$$j_0 +\cdots +j_N-j_\infty=M\no(6.1)$$
for some non-negative integer $M$.
For each $a=1,\cdots,N$ let $\pi_a:\U\rightarrow V_{j_a}(z)$ be the evaluation
representation and $\tilde \Phi_{l,j_a}^m(z)\,:\,\F_l\rightarrow\F_{m}\otimes
V_{j_a}(z)$ be an intertwining operator.
Put
$$\Phi_{l,j_a}^m(z)=z^{\Delta_{m}-\Delta_{l}}\tilde \Phi_{l,j_a}^m(z).$$
In this section we shall be concerned with a formal calculation of the
$n$-point correlation functions:
$$\<\Phi_{l_{N-1}j_N}^{j_\infty}(z_N)\Phi_{l_{N-2},j_{N-1}}^{l_{N-1}}(z_{N-1})
 \cdots \Phi_{l_1,j_2}^{l_2}(z_2)\Phi_{j_0,j_1}^{l_1}(z_1)\>\no(6.2)$$
for a choice of $l_1, \cdots,l_N$.
Here $\<A\>=(\<j_\infty|,A|j_0\>)$ is given by the canonical pairing
$\F_{j_\infty,j_\infty,j_\infty}^{\dagger}\times
\F_{j_\infty,j_\infty,j_\infty}\rightarrow \C$.
\par
However, since the operator $\tilde \Phi_{l,j}^m(z)$ for $m\ne l +j$ is not
defined rigorously at present, we shall use the following function instead:
$$\eqalign{
&\FF(t_1, \cdots,t_M,z_1, \cdots,z_N)\cr
&\qqquad=\<S(t_1) \cdots S(t_M)\Phi_{l_{N-1},j_{N}}^{j_\infty
+M}(z_N)\Phi_{l_{N-2},j_{N-1}}^{l_{N-1}}(z_{N-1}) \cdots
\Phi_{l_1,j_2}^{l_2}(z_2)\Phi_{j_0,j_1}^{l_1}(z_1)\>\cr}\no(6.3)$$
where $l_a=j_0 + \cdots +j_a$.
\par
Let ${\cal R}\in \U\hat\otimes\U$ be the universal $R$-matrix and put
$$R_{V_aV_b}(z_a/z_b)=(\pi_{a}\otimes\pi_{b})\bigl(\sigma({\cal
R}^{-1})\bigr)$$
where $\sigma(u\otimes v)=v\otimes u$ (cf. [FR]).
It acts on the $a$-th and $b$-th components of $V_{j_1}\otimes  \cdots \otimes
V_{j_N}$ as a formal power series of $z_a/z_b$.
Let $T_a f(z_1, \cdots,z_N)=f(z_1, \cdots,pt_a, \cdots,z_N)$.
Then we have the following:
\medn
{\bf Proposition 6.1 }
The function $\FF(t,z)=\FF(t_1, \cdots,t_M,z_1, \cdots,z_N)$ valued in
$V_{j_1}\otimes  \cdots \otimes V_{j_N}$ satisfies the quantum ($q$-deformed)
Knizhnik-Zamolodchikov equation:
$$\eqalign{T_a\FF(t,z)&=R_{V_aV_{a-1}}\bigl({p{z_a} \over
{z_{a-1}}}\bigr) \cdots R_{V_aV_1}\bigl({{pz_a} \over {z_{1}}}\bigr)\cr
 &\qqquad \pi_a(q^{h_1})^{j_0 +j_\infty +1}R_{V_N V_a}\bigl({{z_N} \over
{z_a}}\bigr)^{-1} \cdots R_{V_{a +1} V_a}\bigl({{z_{a +1}} \over
{z_a}}\bigr)^{-1}\FF(t,z),\cr
 &\qquad a=1, \cdots,N,\cr}$$
modulo total $p$-difference of a function with respect to $t_1, \cdots,t_M$,
where $p=q^{2(k +2)}$ as before.
\medn
{\it Proof. }
It is proved by the same way as [FR], [J2] or [IIJMNT] if we note the
following: \par
(1) The Drinfeld Casimir operator acts on $F_l$ by a scalar $p^{\Delta_l}$.\par
(2) The operator $S(t)$ commutes with the action of $\U'$ modulo a total
difference.\parn
Here (1) for general $l$ follows from Corollary 4.4 and the result is continued
to arbitrary $l$, and (2) is nothing else but Proposition 4.5.
\QED
\med
Now let us explicitly calculate the correlation function when the number of the
screening operator is one.
Let us prepare the following notations:
$$(x;p)_\infty=\prod_{m=0}^{\infty}(1-p^mx),\quad
(x;p,q^4)_\infty=\prod_{m_1=0}^{\infty}
\prod_{m_2=0}^{\infty}(1-p^{m_1}q^{4m_2}x).$$
At first, by operator product expansion, we have
$$\eqalign{
&\<S(t)\phi_{j_N,j_N}(z_N) \cdots\phi_{j_{a +1},j_{a +1}}(z_{a
+1})X^-(x)\phi_{j_a,j_a}(z_a) \cdots\phi_{j_1,j_1}(z_1)\>\cr
&\qqquad=D(t,z_1, \cdots,z_N)\psi_a(t,x,z_1, \cdots,z_N),\cr}$$
where
$$\eqalign{
&D(t,z_1, \cdots,z_N)=\prod_{a=0}^Nt^{-{{2j_a} \over{k
+2}}}\prod_{a=1}^N{{(pq^{2j_a}t/z_a;p)_\infty}
\over{(pq^{-2j_a}t/z_a;p)_\infty}}
\prod_{0\leq a<b\leq N}z_b^{{{2j_aj_b} \over{k +2}}}
\cr
&\quad\prod_{1\leq a<b\leq N}
{{(pq^{2j_a +2j_b +2}z_b/z_a;p,q^4)_\infty(pq^{-2j_a-2j_b
+2}z_b/z_a;p,q^4)_\infty} \over{(pq^{2j_a-2j_b
+2}z_b/z_a;p,q^4)_\infty(pq^{-2j_a +2j_b +2}z_b/z_a;p,q^4)_\infty}},\cr}
$$
and
$$\eqalign{
&\psi_a(t,x,z_1, \cdots,z_N)={1 \over{(q-q^{-1})^2tx}}\cr
&\qquad\times\biggl\{q^{-1}\prod_{m=1}^a{{q^{j_m}x-q^{k +2}z_m} \over{x-q^{k +2
+j_m}z_m}}
-{{q^{-1}t-q^{-k-1}x} \over{t-q^{-k-2}x/t}}\prod_{m=1}^aq^{-j_m}\prod_{m=a
+1}^N{{z_m-q^{-k-2-j_m}x} \over{z_m-q^{-k-2 +j_m}x}}\cr
&\qquad\qquad
-{{qt-q^{k +1}x} \over{t-q^{k +2}x}}\prod_{m=1}^i{{q^{j_m}x-q^{k +2}z_m}
\over{x-q^{k +2 +j_m}z_m}}
 +q\prod_{m=1}^aq^{-j_m}\prod_{m=a +1}^N{{z_m-q^{-k-2-j_m}x} \over{z_m-q^{-k-2
+j_m}x}}\biggr\}\cr
}$$
Here each term of $\psi_a(t,x,z_1, \cdots,z_N)$ is understood to be a power
series, by expanding as
$${{u-q^av} \over {u-q^bv}}=(1-q^av/u)\sum_{m=0}^\infty
(q^bv/u)^m\quad (|q^bv/u|<1),$$
which is analytically continued.
\par
Let $\ds{\oint {{dx} \over{2\pi\sqrt{-1}}}}$ denote the integration around
$x=0$ according to the expansion above, and $\Res_{x=x_0}$ denote the residue
at $x=x_0$ of a rational function after continued.
Since $\Res_{x=0}\psi_a(t,x,z_1, \cdots,z_N)=0$, we have
$$\eqalign{
&\oint {{dx} \over{2\pi\sqrt{-1}}}\psi_a(t,x,z_1, \cdots,z_N)\cr
&\qquad={1 \over{(q-q^{-1})^2}}\oint {{dx} \over{2\pi\sqrt{-1}}}{1
\over{tx}}\left\{q^{-1}-{{qt-q^{k +1}x} \over{t-q^{k
+2}x}}\right\}\prod_{m=1}^i{{q^{j_m}x-q^{k +2}z_m} \over{x-q^{k +2
+j_m}z_m}}\cr
&\qquad=-{1 \over{(q-q^{-1})}}\oint {{dx} \over{2\pi\sqrt{-1}}}\,{1
\over{(t-q^{k +2}x)x}}\prod_{m=1}^a{{q^{j_m}x-q^{k +2}z_m} \over{x-q^{k +2
+j_m}z_m}}\cr}$$
Now $\Res_{x=\infty}\psi_a(t,x,z_1, \cdots,z_N)=0$, and
$$\Res_{x=q^{-k-2}t}\psi_a(t,x,z_1, \cdots,z_N)=-{1
\over{(q-q^{-1})t}}\prod_{m=1}^a{{q^{-k-2 +j_m}t-q^{k +2}z_m}
\over{q^{-k-2}t-q^{k +2 +j_m}z_m}}$$
Therefore, by the residue theorem, we have
$$\oint {{dx} \over{2\pi\sqrt{-1}}}\psi_a(t,x,z_1, \cdots,z_N)={1
\over{(q-q^{-1})t}}\prod_{m=1}^a{{q^{j_m}t-q^{2(k +2)}z_m} \over{t-q^{2(k +2)
+j_m}z_m}}$$
Hence we conclude
$$\eqalign{
&\< S(t)\phi_{j_N,j_N}(z_N) \cdots
\phi_{j_a-1,j_a}(z_a) \cdots\phi_{j_1,j_1}(z_1)\>\cr
&\qquad=D(t,z_1, \cdots,z_N)\oint {{dx}
\over{2\pi\sqrt{-1}}}\left\{\psi_{a-1}(t,x,z_1,
\cdots,z_N)-q^{j_a}\psi_a(t,x,z_1, \cdots,z_N)\right\}\cr
&\qquad=D(t,z_1, \cdots,z_N){1 \over
{(q-q^{-1})t}}\left\{1-q^{j_a}{{q^{j_a}t-q^{2(k +2)}z_a} \over{t-q^{2(k +2)
+j_a}z_a}}\right\}\prod_{m=1}^{a-1}{{q^{j_m}t-q^{2(k +2)}z_m} \over{t-q^{2(k
+2) +j_m}z_m}}\cr
&\qquad=D(t,z_1, \cdots,z_N){{(1-q^{2j_a})} \over{q-q^{-1}}}{1 \over{t-q^{2(k
+2) +j_a}z_a}}\prod_{m=1}^{a-1}{{q^{j_m}t-q^{2(k +2)}z_m} \over{t-q^{2(k +2)
+j_m}z_m}}\cr}$$
The correlation function is given by a Jackson integral of this function, which
gives rise to a solution to the quantum Knizhnik-Zamolodchikov equation.
This expression of the integrand is similar to the one considered previously
by the author [M1] up to normalization and transformation.

\beginsection{7. Conclusion}\par
In this paper, we have construcred a $q$-deformtion of the screening operators
and primary fields acting on Fock modules $\F_{l,m_1,m_2}$.
The screening operator $S^+(z)$ is used to define the small Fock space
$\F_l=\ker\ Q^+$, which is same as the Fock space of the Wakimoto module.
The commutater of the screening operator $S(z)$ with $\U'$ is a total
difference, which is eliminated by a Jackson integral in the sense of sect.2.
Thus the correlation function of our primary fields and screening operators
satisfy the quantum ($q$-deformed) Knizhnik-Zamolodchikov equation.
When the number of the screening operator is one, the expression of the
correlation function essentially coincides with the one introduced in [M1].
In this simple case, the calculation is easy because only the primary
fields of type $\phi_{j,j}(z)$ or $\phi_{j,j-1}(z)$ are necessary.
In principle this calculation is generalized to the case that the number of
screening operator is larger than one.
It will be discussed elsewhere.
It is also desirable to study the case of $2j\in\Z_{\geq 0}$, especialy for an
integral level.
It is worth the mention that, to construct the $q$-vertex operators of
irreducible $\U$-modules, the BRST analysis of Bernard-Felder [BF] seems to be
necessary.
\par
Finally we shall make a comment on the work by Kato et al.\ [KQS], which also
gives an expression of the $q$-vertex operators using the Fock module
constructed by Shiraishi [S].
Their preprint seems to contain a serious confusion in the following sense.
The primary fields are a priori operators among the large Fock modules.
In their formulation the small Fock spaces are by defintion $\F_l=\U|l\>$.
This is unsatisfactory since it is not obvious that the primary field is
an operator among them; it could fail when the spin $l$ is special.
Moreover, since the Fock module is reducible, the existence and uniqueness of
the $q$-vertex operators have not been clarified.
It is not certain whether the property (3.16)--(3.18) in [KQS] uniquely
determine the expression of the primary field.
To see it one must know the structure of the Fock module embedded in the large
Fock space in some detail.
Consequently their construction of the $q$-vertex operators seems to be
incomplete.
In the present paper we have overcome these difficulties by introducing the
screening operator $S^+(z)$.
The situation is similar to the $q=1$ case (cf. [FLMSS]).
\par
\beginsection{Acknowledgement.}\par
The author is grateful to K.\ Kimura and H.\ Konno for stimulating
discussion in the early stage of this work.
He also thanks M.\ Kashiwara, M.\ Jimbo and Ya.\ Yamada for
discussion.
He is specialy grateful to A.Tsuchiya for examining the calculation
and suggesting improvements.

\beginsection{References.}\par
\medn
\item{[ABG]} Abada,\ A.,\ Bougourzi,\ A.H.,\ Gradechi,\ M.A.El: Deformation of
the Wakimoto construction, Preprint (1992)
\item{[ATY]} Awata,\ H.,\ Tsuchiya,\ A.,\ Yamada,\ Y.: Integral formulas for
the WZNW correlation functions.\ Nucl.\ Phys.\ B365,\ 680\ (1991)
\item{[BF]} Bernard,\ D.,\ Felder,\ G.: Fock representations and BRST
cohomology in $SL(2)$ current algebra.\ Commun.\ Math.\ Phys.\ 127,\ 145\
(1990)
\item{[B]} Bilal,\ A.: Bosonization of $Z_N$ parafermions and $su(2)_N$
Kac-Moody algebra.\ Phys.\ Lett.\ B226,\ 272\ (1989)
\item{[DJMM]} Date,\ E.,\ Jimbo,\ M.,\ Matsuo,\ A.,\ Miwa,\ T.: Hypergeometric
type integrals and the $sl(2,\C)$ Knizhnik-Zamolodchikov equations.\ Intern.\
J.\ Mod.\ Phys.\ B4,\ 1049\ (1990)
\item{[DFJMN]} Davies,\ B.,\ Foda,\ O.,\ Jimbo,\ M.,\ Miwa,\ T.,\ Nakayashiki,\
A.: Diagonalization of the XXZ Hamiltonian by vertex operators.\ Preprint
RIMS-873 (1992)
\item{[D1]} Drinfeld,\ V.G.: Hopf algebras and the quantum Yang-Baxter
equation.\ Sov.\ Math.\ Dokl.\ 32,\ 254\ (1985).
\item{[D2]} Drinfeld,\ V.G.: A new realization of Yangians and quantum affine
algebras.\ Sov.\ Math.\ Dokl.\ 36,\ 212\ (1988).
\item{[FLMSS]} Frau,\ M.,\ Lerda,\ A.,\ McCarthy,\ J.G.,\ Sciuto,\ S.,\
Sidenius,\ J.: Free field representation for $\widehat {SU(2)}_k$ WZNW models
on Riemann surface.\ Phys.\ Lett.\ B 245,\ 453\ (1990)
\item{[FJ]} Frenkel,\ I.B.,\ Jing,\ N.H.: Vertex representations of quantum
affine algebras.\ Proc.\ Nat'l\ Acad.\ Sci.\ USA\ 85,\ 9373\ (1988)
\item{[FR]} Frenkel,\ I.B.,\ Reshetikhin,\ N.Yu.: Quantum affine algebras and
holonomic difference equations.\ Commun.\ Math.\ Phys.\ 146,\ 1\ (1992)
\item{[FMS]} Friedan,\ D.,\ Martinec,\ E.,\ Shenker,\ S.: Conformal invariance,
supersymmetry and string theory.\ Nucl.\ Phys.\ B271,\ 93\ (1986)
\item{[IIJMNT]} Idzumi,\ M.\ Iohara,\ K.\ Jimbo,\ M.,\ Miwa,\ T.,\ Nakashima,\
T.,\ Tokihiro,\ T.: Quantum affine symmetry in vertex models.\ Preprint (1992)
\item{[I]} Ito,\ K.: $N=2$ super coulomb gas formalism.\ Nucl.\ Phys.\ B332,\
566\ (1990)
\item{[JNS]} Jayaraman,\ T.,\ Narain,\ K.S.,\ Sarmadi,\ M.H.: $SU(2)_k$ WZW and
$Z_k$ parafermion models on the torus.\ Nucl.\ Phys.\ B343,\ 418\ (1990)
\item{[J1]} Jimbo,\ M.: A $q$-difference analogue of U($\ge$) and the
Yang-Baxter equation.\ Lett.\ Math.\ Phys.\ 10,\ 63\ (1985)
\item{[J2]} Jimbo,\ M., lecture at Tokyo university.
\item{[JMMN]} Jimbo,\ M.,\ Miki,\ K.,\ Miwa,\ T.,\ Nakayashiki,\ A.:
Correlation functions of the XXZ model for $\Delta<-1$.\ Preprint\ RIMS-877\
(1992)
\item{[KZ]} Knizhnik,\ V.G.,\ Zamolodchikov,\ A.B.: Current algebra and
Wess-Zumino models in two dimensions.\ Nucl.\ Phys.\ B247,\ 83\ (1984)
\item{[KQS]} Kato,\ A.,\ Quano,\ Y.,\ Shiraishi,J.: Free boson representation
of $q$-vertex operators and their correlation functions.\ Preprint\ UT-618\
(1992)
\item{[L]} Lusztig,\ G.: Quantum deformation of certain simple modules over
enveloping algebras.\ Adv.\ Math.\ 70,\ 237\ (1988)
\item{[M1]} Matsuo,\ A.: Jackson integrals of Jordan-Pochhammer type and
quantum Knizhnik-Zamolodchikov equations.\ To appear in Commun.\ Math.\ Phys.
\item{[M2]} Matsuo,\ A.: Quatum algebra structure of certain Jackson
integrals.\ Preprint\ (1992)
\item{[M3]} Matsuo,\ A.: Free field representation of the quantum affine
algebra $U_q(\widehat{\sl}_2)$.\ Preprint\ (1992)
\item{[N]} Nemeschansky,\ D.: Feigin-Fuchs representation of
$\widehat{su}(2)_k$ Kac-Moody algebra.\ Phys.\ Lett.\ B224,\ 121\ (1989)
\item{[R]} Reshetikhin,\ N.Yu.: Jackson type integrals, Bethe vectors, and
solutions to a difference analog of the Knizhnik-Zamolodchikov system.\
Preprint\ (1992)
\item{[SV]} Schechtman,\ V.V., Varchenko,\ A.N.: Arrangements of hyperplanes
and Lie algebra homology.\ Invent.\ Math.\ 106,\ 139\ (1991)
\item{[S]} Shiraishi,\ J.: Free boson representation of $U_q(\hat {\sl}_2)$.\
Preprint\ UT-617\ (1992)
\item{[TK]} Tsuchiya,\ A.,\ Kanie,\ K.: Vertex operators in conformal field
theory on ${\bf P}^1$ and monodromy representations of braid groups.\ Adv.\
Stud.\ Pure.\ Math.\ 16,\ 297\ (1988)
\item{[W]} Wakimoto,\ M.: Fock representation of the affine Lie algebra
$A_1^{(1)}$.\ Commun.\ Math.\ Phys.\ 104,\ 605\ (1986)
\bye